\documentclass{FM6_Focus_in_Astronomy}
\usepackage{graphicx}

\title[AM Evolution in $\Lambda$CDM] 
{The Fundamental Physics of Angular Momentum Evolution in a $\Lambda$CDM Scenario}

\author[Susana Pedrosa]   
{Susana Pedrosa$^1$ $^2$}

\affiliation{$^1$Institute for Astronomy and Space Science, CONICET - UBA, \\ Ciudad Universitaria,
Buenos Aires, Argentina \\ email: {\tt supe@iafe.uba.ar} \\[\affilskip]
$^2$Dept. de F\'{\i}sica Te\'orica, Univ. Aut\'onoma de Madrid\\ Cantoblanco, Madrid, Espana \\}

\pubyear{2018}
\jname{Galactic Angular Momentum, Volume 1} 
\editors{Patricia Tissera \& Danail Obreschkow, ed.}
\begin{document}

\maketitle

\begin{abstract}

Galaxy formation is a very complex process in which many different physical mechanisms intervene. Within the LCDM paradigm processes such as gas inflows and outflows, mergers and interactions contribute to the redistribution of the angular momentum content of the structures. Recent observational results have brought new insights and also triggered several theoretical studies. Some of these new contributions will be analysed here. 

\keywords{galaxies: formation, galaxies: evolution}

\end{abstract}

\firstsection 
\section{Introduction}
 
Angular momentum exchange is ubiquitous in the structure assembly process. Every galaxy formation scenario assumes the exchange of angular momentum. Encrypted inside the today galactic morphology is stored the assembly history not only of the galaxy itself but also of the different regions of the DM halo. Estimations of the angular momentum content of the dark matter haloes require a connection between it and the galactic one. 

In a hierarchical clustering universe the angular momentum budget of galaxies originates in primordial torques that act upon baryons and the dark matter. \cite[Fall (1983)]{Fall83} proposed, in a seminal paper, the fundamental correlation between the angular moment (AM) content of the galactic components and the stellar mass of the galaxy. He found that both components follow a power law correlation with the stellar mass, with an exponent of $\sim 0.6$. The spheroidal component, although following a parallel sequence, presents an offset a factor of about 5 lower due to the lost of AM during the galaxy assembly. Theoretical models of galaxy formation (\cite[Fall \& Efstathiou 1980]{FE80}, \cite[Mo et al. 1998]{Mo98}) predict a linear correlation between the dark matter specific angular momentum and $M_{virial}^{2/3}$.

Within the hierarchical scenario galaxies are shaped and reshaped by several processes. For instance, supernova feedback that redistributes energy and mass through mass loaded winds is the key ingredient in the galaxy formation recipe that allows theoretical models to overcome the "angular momentum catastrophe€. With its inclusion more realistic galaxies could be obtained in cosmological simulations. Other processes that may ultimately determine the resulting galactic morphology are mergers, interactions and disc instabilities. For instance, mergers, in all their types, are the most accepted mechanism responsible for the formation of spheroidal galaxies.

\section{Observational studies} 

In 2012 and 2013, \cite[Fall \& Romanowsky (2013)]{FR13} (FR13) and \cite[Romanowsky \& Fall (2012)]{FR12}) (FR12), revisited \cite[Fall (1983)]{Fall83} using an improved and extended observational sample. They confirmed previous findings: all morphological types of galaxies lie along a parallel sequence with exponent $\alpha \sim 0.6$ in the stellar j - M plane. They proposed then that the j - M diagram constitutes a more physically motivated description of the galactic morphology than the typical disk to bulge classification.

\cite[Obreschkow \& Glazebrook (2014)]{OG14}, using high precision measurements of the stellar and baryonic specific AM of a THINGS sample of 16 spiral galaxies (\cite[Leroy et al. 2008]{Leroy08}), includes the $\beta$ parameter (bulge fraction) in the AM description. They found a strong correlation for the plane fitting the 3D space of $\beta$, logM and log j. For a fixed $\beta$, the projection results in an exponent $\alpha \sim 1$, larger than the one found by \cite[Romanowsky \& Fall (2012)]{FR12}. They proposed that the contributions to the j-M plane of bulge and disks components were not independent.

After the FR12 and FR13 works, it followed a most interesting burst of numerical studies analysing to what extent their models fitted these observational constrains. Interestingly, despite the fact that this numerical experiments were perform with different numerical codes, different  prescriptions for the subgrid physics, different resolutions, different feedback implementations, they all results in very good agreement with the new observational data. And through this process, many interesting contributions to the knowledge about the origin and evolution of the galactic AM during the assembly of the galaxy were developed. Numerical simulations constitute ideal tools for filling the gap between the observed and the inferred knowledge.

\begin{figure}[b]
\begin{center}
 \includegraphics[width=3.8in]{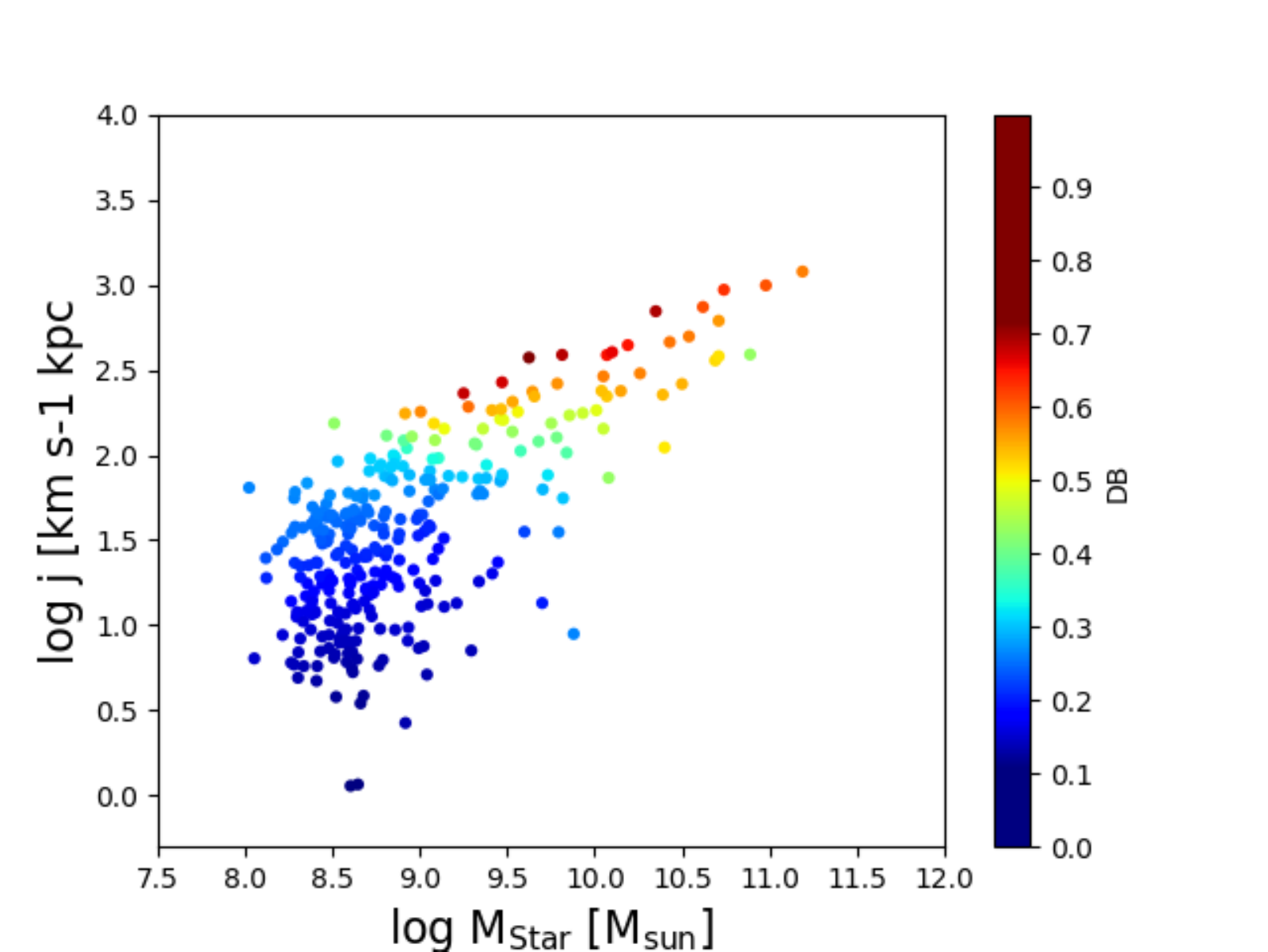} 
 \caption{Relation between j and the total stellar mass for simulated galaxies with different disc-to-bulge ratios}
   \label{fig2}
\end{center}
\end{figure}

\section{Numerical analysis}

\cite[Genel et al. (2015)]{Genel15} analysed the AM distribution of galaxies from the Illustris Simulation (\cite[Vogelsberger et al. 2014]{Vo14}, \cite[Genel et al. 2014]{Genel14}). They discriminate the $z=0$ population based on their specific star formation rate, the flatness and the concentration, obtaining two parallel relations corresponding to early-and-late-type galaxies, in agreement with FR12 observations. They find that galactic winds with high mass-loading factors are crucial for getting the late-type galaxies relation that results from full conservation of the specific AM generated by cosmological tidal torques. 

Using intermediate resolution cosmological simulations, \cite[Pedrosa \& Tissera (2015)]{Pedrosa15} found the specific angular momentum of spheroidal and disk component to determine relations with the same slopes, regardless of the virial mass of their host galaxy. They found no evolution of this relation with redshift, indicating that spheroidal and disk component conserve similar relative amount of AM as they evolve,  independently of virial masses. As shown in Fig.\ref{fig2}, there is a clear correlation between the morphological type of the galaxy and its total specific AM content: higher D/T ratios are related with higher contents of specific AM. The AM of stellar bulges is consistent with elliptical galaxies indicating that bulges might be considered as mini-ellipticals, in agreement with FR12.

\begin{figure}
\begin{center}
 \includegraphics[width=3.5in,trim={16mm 50mm 30mm 55mm},clip]{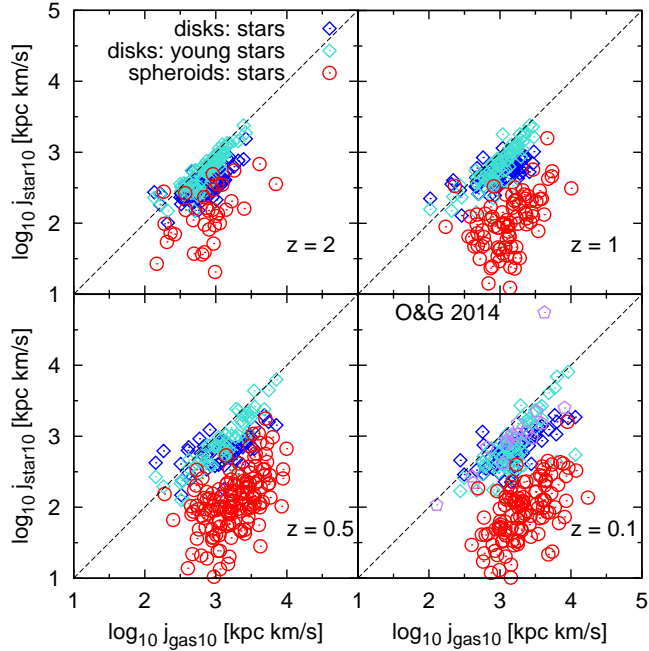} 
 \caption{[Reprint from \cite[Teklu et al. (2015)]{Teklu15}] Specific angular momentum of the gas against the specific angular momentum of stars, both within 10\% of the virial radius for galaxies that are classified as disks (blue diamonds) at four redshifts. For young stars only (turquoise diamonds). At z = 0.1 comparison with data from \cite[Obreschkow \& Glazebrook (2014)]{OG14} (purple pentagons).}
   \label{fig3}
\end{center}
\end{figure}

\begin{figure}[h]
\begin{center}
 \includegraphics[width=3in,trim={20mm 30mm 12mm 55mm},clip]{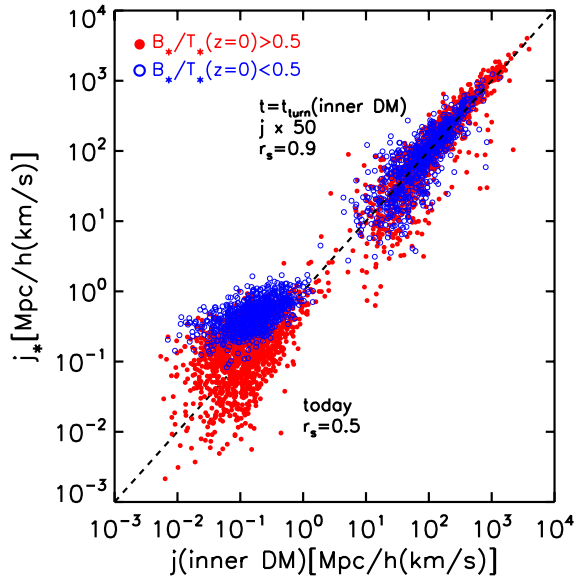} 
 \caption{[Reprint from \cite[Zavala et al. (2016)]{Zavala16}] Correlation between the stellar specific angular momentum and that of the inner dark matter halo at $z = 0$ and at the time of turnaround of the inner dark matter halo (for the latter epoch, j is multiplied by 50). The sample of galaxies is divided into present-day bulge-dominated (solid red), and disc-dominated (open blue).}
   \label{fig4}
\end{center}
\end{figure}

\cite[Teklu et al. (2015)]{Teklu15}, using galaxies from the Magneticum Pathfinder Project (Dolag et al. 2015, in preparation), find that disk galaxies populate haloes with larger spin than those that host spheroidal galaxies. And disk galaxies live preferentially in haloes with central AM aligned with the AM of the whole halo. They also verify that their galaxies are located on the j-M plane in agreement with observations. Their stellar disk AM is lower than the cold gas one, in agreement with \cite[Pedrosa et al. (2015)]{Pedrosa15} and \cite[Obreschkow \& Glazebrook (2014)]{OG14}. They attribute this excess to the recent accretion of gas with high AM from the outer parts of the halo. This is shown in the fact that young stars (formed from this freshly accreted gas) present higher content of AM than older ones, Fig.~\ref{fig3}.

\cite[Zavala et al. (2016)]{Zavala16}, using a sample of over 2000 central galaxies extracted from the EAGLE simulation (\cite[Schaye et al. 2015]{Schaye15}, \cite[Crain et al. 2015]{Crain15}) follow through time selected particles at z = 0 using a Lagrangian method. They find a correlation between the specific AM of $z=0$ stars and that of the inner part of the DM halo. They find this link to be specially strong for stars formed before the turnaround. Spheroids, typically assembled at this epochs will then suffer loss of AM due to the merging activity of the inner halo assembly process. The cold gas, that mostly preserves the high specific AM acquired from the primordial tidal torques, forms stars after the turnaround and then build the stellar disk component. They find that the inner DM halo loses $90\%$ of its specific AM after turnaround through transfer to dark matter clumps. While bulge dominated objects tracks the inner halo, the disk dominated ones follow the whole DM halo closely, Fig.~\ref{fig4}. They claim that most of the stars belonging to the today ellipticals were already formed at turnaround and then got locked inside the DM clumps that will form the inner halo. 

\begin{figure}
\begin{center}
 \includegraphics[width=\textwidth]{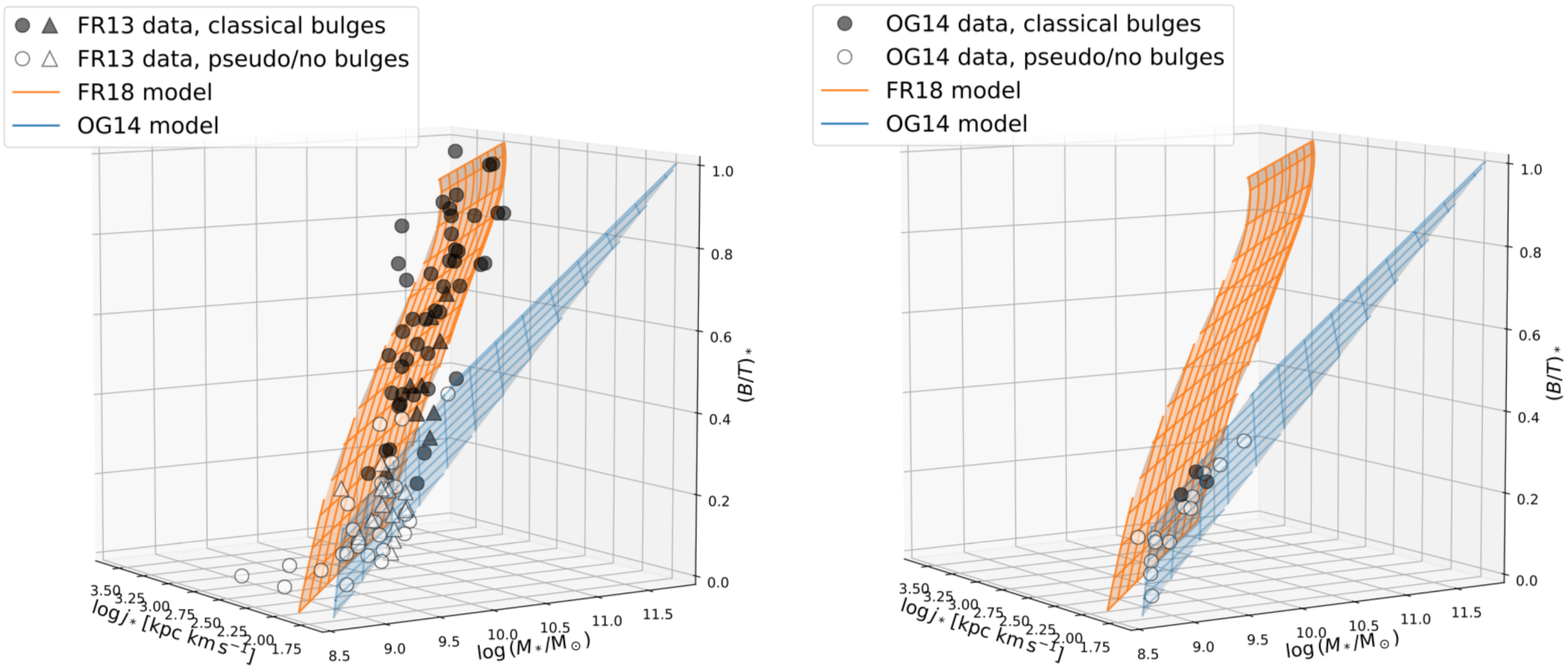} 
 \caption{[Reprint from Fall \& Romanowsky (2018), arXiv:1808.02525] Stellar bulge fraction beta against stellar specific angular momentum j and stellar mass M from Fall \& Romanowsky (2018) (left panel) and from \cite[Obreschkow \& Glazebrook (2014)]{OG14} (right panel). Filled symbols: galaxies with classical bulges, open symbols:pseudo bulges or no bulges. Orange surface:relation for independent disks and bulges derived in Fall \& Romanowsky (2018), while the blue plane: linear regression derived by \cite[Obreschkow \& Glazebrook (2014)]{OG14}.}
   \label{fig6}
\end{center}
\end{figure}

Also using a Lagrangian method, \cite[Obreja et al (2018)]{Obreja18} trace back in time structure progenitors of 25 zoom-ins simulations from the NIHAO Project (\cite[Wang et al. 2015]{Wang15}), in order to dissect the AM budget evolution of the eight morphological components they identify (see also \cite[Dominguez Tenreiro et al (2015)]{Domin15} . They find that thin disks typically retain 70\% of its AM while thick disks only a 40\%. They also find that 90\% of their velocity dispersion dominated objects in the sample retain less than 10\% of the central AM. Regarding the rotation dominated structures, most of the thin disks has a retention factor greater than 50\% while thick disks might loose as much as 85\% of its AM.

Most recently,\cite[Fall \& Romanowsky (2018)]{FR18} presented new observational evidence that reinforce their previous assumptions. They propose a simple model, valid for all morphological type, in which the stellar AM is the linear superposition of independent contributions from disks and bulges. They obey a power scaling relation with essentially the same coefficient but differs in their normalisation. They consider that the parallel sequences in the j - M plane corresponding to different values of $\beta$ (bulge fraction) means that disks and spheroids are formed via independent processes. Spheroidal galaxies never acquired enough AM or else they loose it in violent events, while disk-dominated systems are the result of a more quiescent processes that were not affected by mergers.
One of the main feature of the new dataset studied in F\&R18 is that both, photometric and kinematic data, extend to large radii, taking into account tha fact that much of the AM lies beyond the effective radius. This issue was already pointed out by \cite[Lagos et al. (2018)]{Lagos18}. Using simulated galaxies from EAGLE, they find that elliptical galaxies with the higher Sersic index have most of their stellar AM budget inhabiting beyond five half mass radius. 
Fall \& Romanowski (2018) claim that the $\beta$ - j - M 3D diagram is well fitted by a plane modeled as a linear superposition of independent contribution from disks and bulges, Fig.~\ref{fig6}. 

\section{Conclusions}

In the last few year substantial progress has been made in our understanding of the physical processes involved in the evolution of the galactic angular momentum throughout the assembly process of the galaxy. A bigger picture has been build that allows us to grasp the morphological classification of galaxies in terms of their positions in the angular momentum - Mass plane and the processes that determine how this position will evolve. But still important questions remains to be solved. For instance, why both, disk dominated and spheroidal galaxies, follow parallel relations independently of the mass. Higher resolution simulation and improved methods for mimic observations would probably bring some answers and, of course, new questions.

\end{document}